\documentclass[twocolumn,showpacs,amsmath,amssymb,pre,aps]{revtex4}   %%  4-pages !!
\usepackage{graphicx}% Include figure files
\usepackage{dcolumn}% Align table columns on decimal point
\usepackage{bm}% bold math
\usepackage{color}% color text

\renewcommand{\vec}[1]{\mathbf{#1}}

\newif\ifgraph

\graphtrue             %
\begin{document}
\title{
Giant Negative Mobility of Janus Particles in a Corrugated Channel}

\author{Pulak K. Ghosh$^{1}$, Peter H\"anggi $^{2,3}$, Fabio Marchesoni$^{4,5}$, and Franco Nori$^{5,6}$}
\affiliation{$^{1}$Department of Chemistry, Presidency University,
Kolkata - 700073, India} \affiliation{$^{2}$Institut f\"ur
Theoretische Physik, Universit\"at Augsburg, D-86135 Augsburg,
Germany} \affiliation{$^{3}$ Center for Phononics and Thermal Energy
Science and School of Physical Science and Engineering, Tongji
University, 200092 Shanghai, People's Republic of China}
\affiliation{$^{4}$Dipartimento di Fisica, Universit\`{a} di
Camerino, I-62032 Camerino, Italy} \affiliation{$^{5}$CEMS, RIKEN,
Saitama, 351-0198, Japan} \affiliation{$^{6}$Physics Department,
University of Michigan, Ann Arbor, MI 48109-1040, USA}

\date{\today}

\begin{abstract}
We numerically simulate the transport of elliptic Janus particles
along narrow two-dimensional channels with reflecting walls. The
self-propulsion velocity of the particle is oriented along either
their major (prolate) or  minor axis (oblate). In smooth channels, we
observe long diffusion transients: ballistic for prolate particles
and zero-diffusion for oblate particles. Placed in a rough channel,
prolate particles tend to drift against an applied drive by tumbling
over the wall protrusions; for appropriate aspect ratios, the modulus
of their negative mobility grows exceedingly large (giant negative
mobility). This suggests that a small external drive suffices to
efficiently direct self-propulsion of rod-like Janus particles in
rough channels.
\end{abstract}
\maketitle

\section{Introduction}

Self-propulsion \cite{Purcell} is the ability of most living
organisms to move, in the absence of external drives, thanks to an
``engine'' of their own. Self-propulsion of fabricated micro- and
nano-particles (artificial microswimmers)\cite{Schweitzer,22,23,24}
is a topic of current interest in view of applications to
nanotechnology. Their direct experimental observation became
affordable with the synthesis of a new class of asymmetric
microswimmers, which propel themselves by generating local gradients
in the suspension environment (self-phoretic effects)
\cite{Sen_propulsion}. Such particles, called Janus particles (JPs)
\cite{Chen,42}, consist of two distinct sides, only one of which is
chemically or physically active. Thanks to their functional
asymmetry, these active particles can induce either concentration
gradients (self-diffusiophoresis) by catalyzing a chemical reaction
on their active surface \cite{Paxton1,Gibbs,Howse,Bechinger}, or
thermal gradients (self-thermophoresis), e.g., by inhomogeneous light
absorption \cite{Sano} or magnetic excitation \cite{ASCNano2013JM}.

\begin{figure}
\centering
\includegraphics[width=0.45\textwidth]{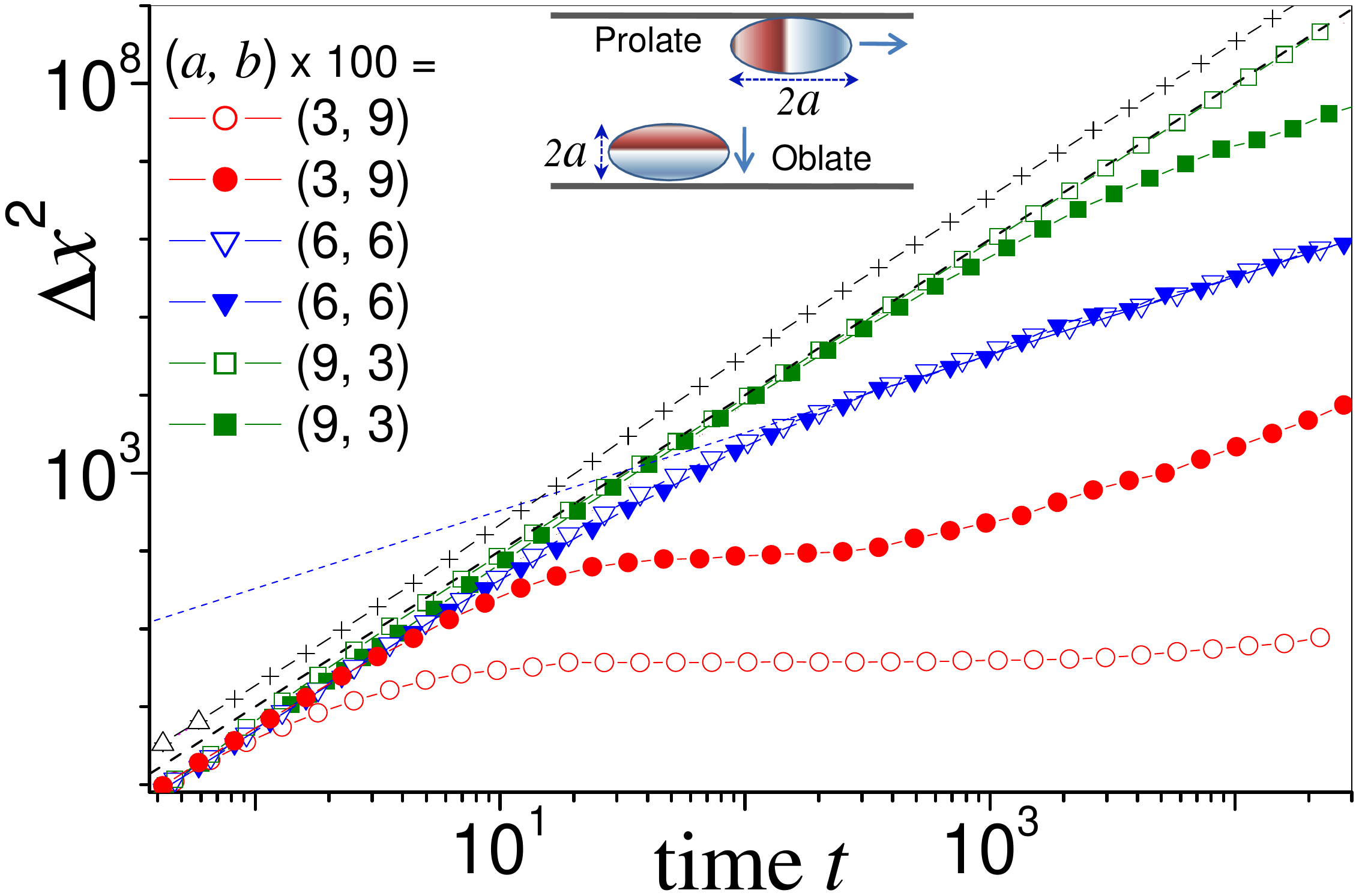}
\caption {(Color online) Diffusion of elliptical JPs with semi-axes
$a$ and $b$ (see legend) in a smooth channel of width $y_L=1$.
Thermal noise is $D_0=10^{-4}$ (empty symbols) and $6.4\times
10^{-3}$ (filled symbols); self-propulsion parameters are $v_0=1$ and
$D_\theta=0.03$. The ballistic transient, $\Delta x^2 = (v_0 t)^2$
(dashed line), and the normal diffusion law, $\Delta x^2 = 2 D_{\rm
eff} t$ (see text, dotted line) are drawn for a comparison. Example
of persistent ballistic diffusion transient (crosses): $a=0.15$,
$b=0.03$, $y_L$=10, $v_0=2$, $D_0=0.02$ and $D_\theta=0.003$
\cite{Bechinger}. Sketch: sliding prolate (top) and stuck oblate disk
(bottom). In both cases $\vec{v}_0$ (single-pointed arrow) is
oriented along the $a$ axis.} \label{F1}
\end{figure}

Much effort is presently directed to achieve reliable transport
control of JPs in confined geometries \cite{Sen_rev,RMP2009,Annalen}.
The ability of Janus microswimmers to perform directed autonomous
motions through periodic arrays \cite{Bechinger} and asymmetric
channels \cite{ourPRL} surely is a suggestive option. Such devices do
operate in the absence of external drives or gradients, but at the
price of strict fabrication requirements regarding their geometry. In
this Letter we propose a more affordable option to direct the motion
of JPs along a channel. Under appropriate conditions involving the
geometry of both the particle and the channel, a tiny external drive
(even in the absence of other biases) can orient the self-propulsion
velocity of the microswimmers {\it against} the drive; a phenomenon
known as absolute negative mobility (ANM) \cite{ANM,ANM2}. The
roughness of the channel walls, mimicked here by randomly inserting
small transverse wall protrusions, can drastically enhance this
phenomenon, thus producing a {\it giant} absolute negative mobility.
%%%%%%%%%%%%%%%%%%%%%%%%%%%%%%%%%%%%%%%%%%%%%%%%%%%%%%%%%%%%%%%%%%%%%%%%%%%%%%%5
%Further
%increasing the external drive induces current inversions (mobility
%zeros) and, therefore, negative differential mobility regimes.
These features suggest most sensitive control techniques on JP
transport with beneficial applications to nanotechnology and medical
sciences \cite{Schweitzer,Chen}

\section{Model}

Shape is known to play a central role in the diffusive dynamics of
confined JPs \cite{Lugli,Gompper}. For this reason we consider
two-dimensional (2D) channels and elongated particles, modelled as
elliptical disks with semi-axes $a$ and $b$. Actual rod-like JPs can
be synthesized through a variety of well-established techniques
\cite{Sen_rev,Lugli,Sen_rod}. An elongated JP  gets a continuous push
from the suspension fluid, which in the overdamped regime amounts to
a self-propulsion velocity ${\vec v_0}$ with constant modulus $v_0$.
Additional dynamical effects, neglected in the present model, are
briefly discussed at the end. We assume that ${\vec v_0}$ acts along
the $a$ axis of the particle: $\vec {v}_0$ is parallel to the
particle major axis for $a>b$ (prolate JP) and orthogonal to it for
$a<b$ (oblate JP). The self-propulsion direction varies randomly with
time constant $\tau_\theta$; accordingly, the microswimmer mean
self-propulsion path is $l_\theta=v_0\tau_\theta$.

The bulk dynamics of such a JP obeys the Langevin equation
\cite{Marchetti}
\begin{equation}
\label{LE} \vec{\dot r}=\vec{v}_0(t)+\vec{F}+\vec {\xi}_0(t),
\end{equation}
where $\vec{r}=(x, y)$ denotes the position of the particle center of
mass, $\vec{F}=(F, 0)$ represents a d.c. external bias parallel to
the channel axis, and $\vec{\xi}_0(t)=(\xi_{0,x}(t), \xi_{0,y}(t))$
models a Gaussian thermal noise with $\langle \xi_{0,i}(t)\rangle=0$
and $\langle \xi_{0,i}(t)\xi_{0,j}(0)\rangle=2D_0\delta_{ij}\delta
(t)$, with $i, j=x,y$, The self-propulsion velocity ${\vec
v_0}=v_0(\cos \theta, \sin \theta)$ is oriented at an angle
$\theta(t)$ with respect to the $x$ axis, where $\theta(t)$ is a
Wiener process, $\dot \theta=\xi_\theta(t)$, with $\langle
\xi_{\theta}(t)\rangle=0$ and $\langle
\xi_{\theta}(t)\xi_{\theta}(0)\rangle=2D_\theta\delta (t)$. As a
change in $\theta$ corresponds to a rotation of the swimmer,
$D_\theta$ is related to both $\tau_\theta$, $D_\theta=2/\tau_\theta$
\cite{Marchetti,ourPRL}, and the average thermal diffusivity $D_0$,
$D_\theta \propto D_0/ab$ \cite{Lowen_basic}. To make contact with
realistic experimental conditions in our simulations we chose
parameters experimentally accessible. For instance, expressing times
in seconds and lengths in microns, typical values for spherical JP of
radius $0.5-2.0$ are $D_0=0.02-0.03$, $D_\theta = 0.01$, and
$v_0=0.05-0.5$ \cite{Bechinger}

When confined to a cavity of size smaller than its self-propulsion
length, $l_\theta$, the microswimmer undergoes multiple collisions
with the walls and the confining geometry comes into play (Knudsen
diffusion \cite{Brenner}). The Langevin equations (\ref{LE}) have
been numerically integrated under the assumption that the channel
walls were perfectly reflecting and the particle-wall collisions
elastic \cite{ANMshape}.

\section{Diffusion transients}

Transport of elongated JPs along a smooth channel is characterized by
unit mobility, $\mu=1$, where $\mu \equiv v/F$ and $v(F)=\lim_{t\to
\infty} \langle x(t)\rangle/t$. The particle shape becomes
distinguishable when one looks at the dispersion, $\Delta
x^2(t)=\langle x^2(t)\rangle -\langle x(t)\rangle^2$. Both prolate
and oblate disks undergo surprisingly long transients, see Fig.
\ref{F1}, before the expected normal diffusion regime sets on. On the
contrary, circular disks seem to diffuse like ordinary point-like JPs
\cite{Marchetti,ourPRL}: for $t\gtrsim \tau_\theta$, $\Delta x^2(t)=
2D_{\rm eff}t$ with $D_{\rm eff}=D_0+v_0l_\theta/4$. Due to their
collision against the walls, active prolate swimmers tend to slide
parallel to the channel axis, which explains their ballistic
diffusion transient with $\Delta x^2(t) \sim (v_0 t)^2$, whereas
oblate swimmers tend to pile up against the walls, thus suppressing
longitudinal diffusion. Upon increasing the strength of the thermal
noise, $D_0$, both transients are shortened, though at a different
rate. This property can be explained by noticing that the onset of
normal diffusion requires that the microswimmers are capable of
inverting the direction of $\vec {v}_0$ against the confining action
of the channel walls. On the other hand, a U-turn can only occur when
noise kicks the elongated particle out of its steady (sliding or
stuck) state. This amounts to a noise-activated process with time
constant of the order of \cite{ANMshape} $\tau_U\simeq
\tau_0\exp(|a-b|v_0/D_0)$, where $\tau_0 \sim \sqrt{v_0a^2/b}$.
Accordingly, for the  $a$ and $b$ values of Fig. \ref{F1}, transients
are longer for the prolate than for the oblate particle. When
$\tau_U$ becomes of the order of or smaller than the crossing time
$y_L/v_0$, boundary effects vanish and (bulk) normal diffusion is
fully restored, i.e., transients become negligible. For strongly
prolate, say, rod-like swimmers, ballistic transients may extend over
exceedingly long time intervals even at room temperature (see example
in Fig. \ref{F1}). For this reason determining how dispersion scales
with time can prove  a delicate experimental issue \cite{Schweitzer}.

\begin{figure}
\centering
\includegraphics[width=0.45\textwidth]{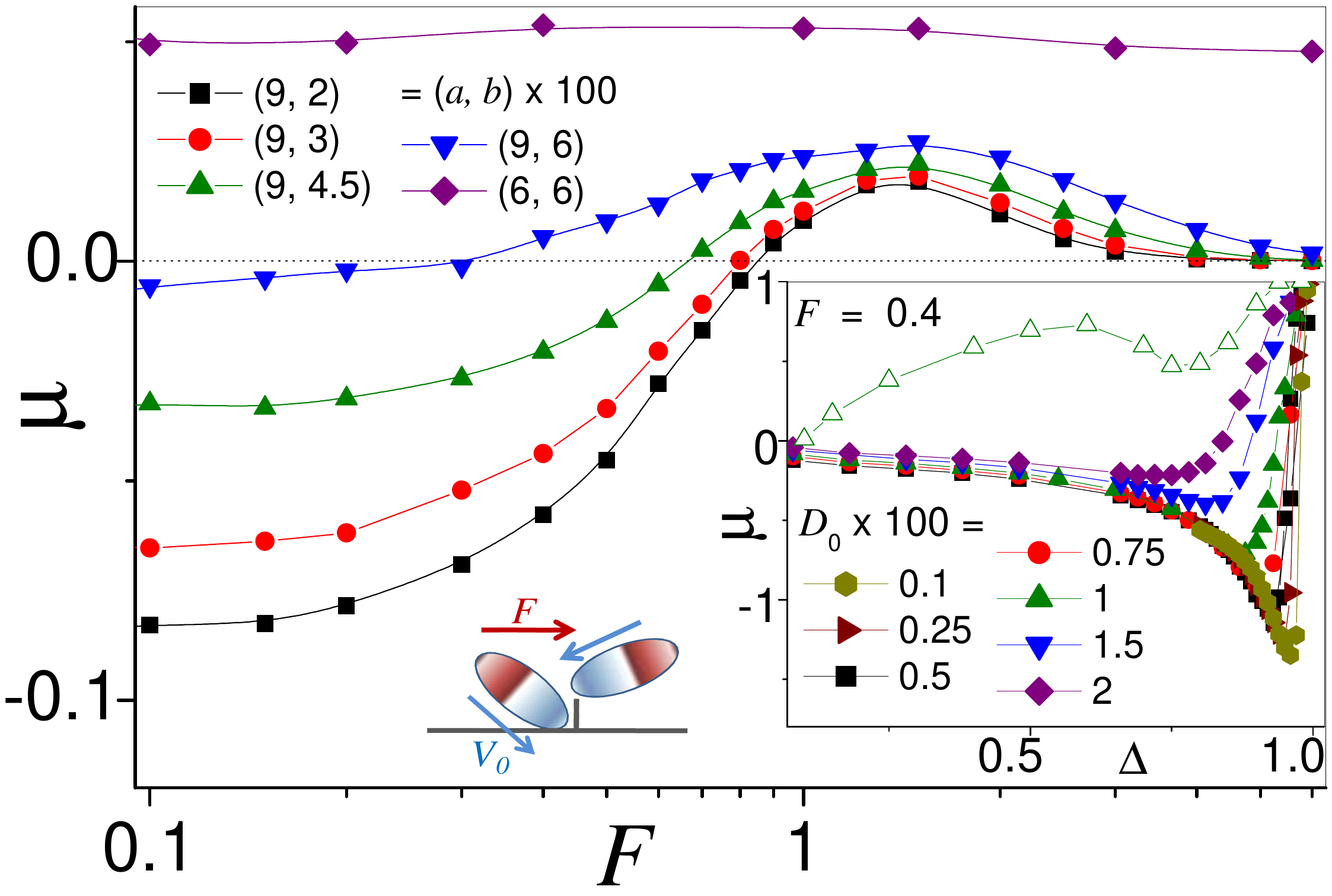}
\caption {(Color online) Mobility $\mu(F)$ of a prolate JP driven
along a septate channel for different values of its semiaxes $a$
and $b$ (see legend). Compartment parameters: $x_L=y_L=1$ and
$\Delta=0.16$; self-propulsion parameters: $v_0=1$ and
$D_\theta=D_0=0.03$. Inset: $\mu$ vs $\Delta$ for $F=0.4$,
$a=0.09$, $b=0.03$ and different $D_0$ (solid symbols). The curve
(empty triangles) for the oblate JP with $a=0.03$, $b=0.09$ at
$D_0=0.01$ is plotted for comparison. All the remaining parameters
are as in the main panel. Sketch: a prolate JP tumbling over a
winglet of the channel wall under the action of the drive (see
text). }\label{F2}
\end{figure}

\section{Absolute negative mobility}

\begin{figure}
\centering
\includegraphics[width=0.45\textwidth]{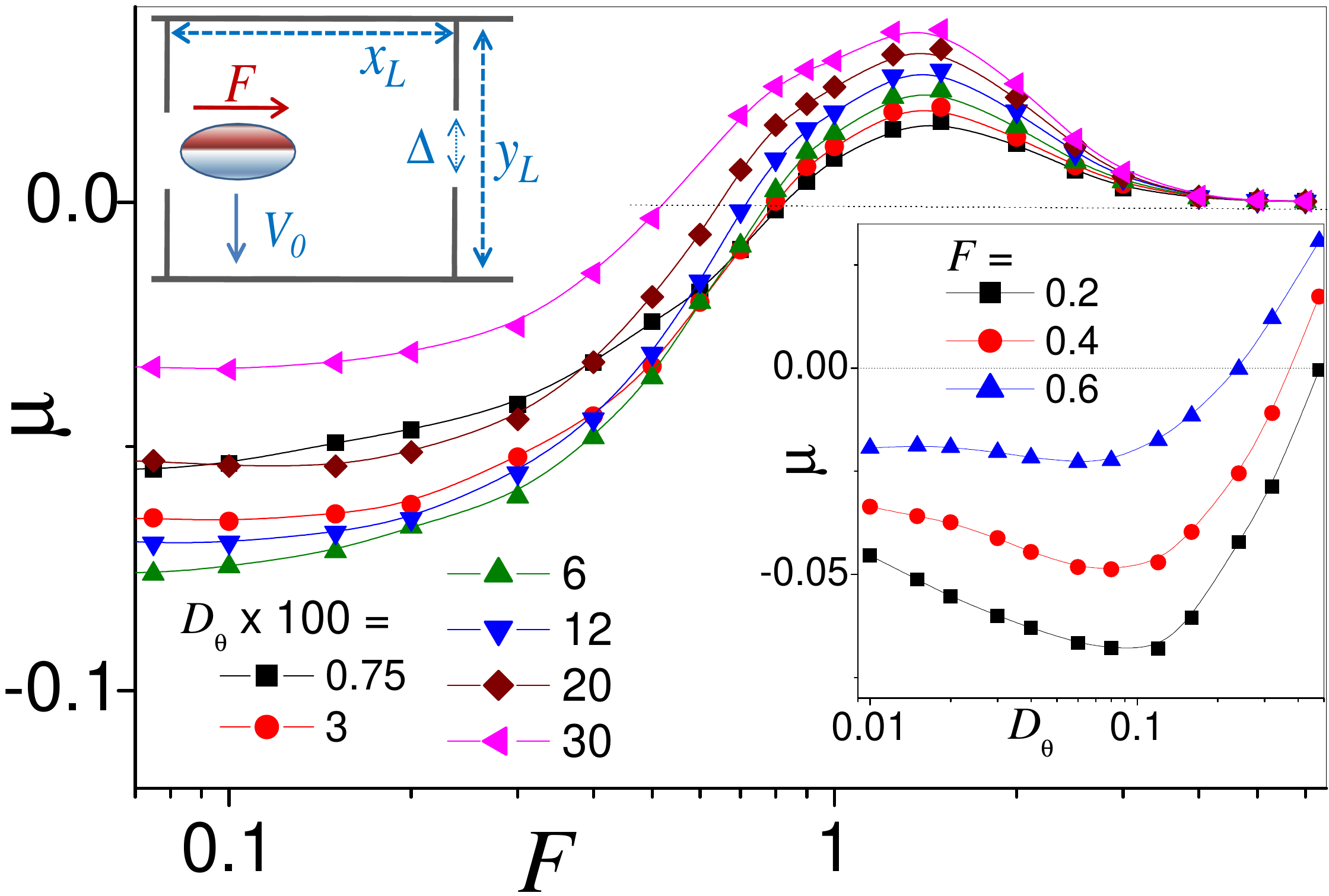}
\caption {(Color online) Mobility $\mu(F)$ of a prolate JP driven
along a septate channel for different values of $D_\theta$ (i.e.,
the reciprocal of $\tau_\theta$, see legend). Compartment
parameters: $x_L=y_L=1$ and $\Delta=0.16$; other simulation
parameters: $v_0=1$, $a=0.09$, $b=0.03$ and $D_0=0.03$. Inset:
$\mu$ vs. $D_\theta$ for different $F$. All remaining parameters
are as in the main panel. Sketch: oblate JP aligned as to leave a
septate compartment to the left, being pulled back by $F$.}
\label{F3}
\end{figure}

We consider now the case of a periodically-compartmentalized channel
obtained by inserting equally-spaced transverse dividers, each
bearing a small opening, or pore, centered on the channel axis. Let
$x_L$ and $y_L$ be the longitudinal and transverse dimensions of the
compartments and $\Delta$ the pore size (see sketch in Fig.
\ref{F3}). Such a channel is termed septate
\cite{septate,septate2,septate3} to stress the role of the
compartment walls when compared with the smooth cross-section
modulation of the so-called entropic channels \cite{ChemPhysChem}.
Here, transport of elongated swimmers is governed by the pore
crossing dynamics, which makes the mobility shape-dependent.

In Figs. \ref{F2} and \ref{F3} we plot the mobility of prolate JPs as
a function of the drive, for different values of the aspect ratio
$a/b$ and the orientation constant $D_\theta$. The most prominent
feature is the appearance of ANM branches at low $F$, which means
that, under certain conditions, active swimmers may drift against the
applied drive \cite{ANM}. This effect (i) grows more apparent for
large aspect ratios, $a/b$ (Fig. \ref{F2}); (ii) is limited to the
domain $F<v_0$; (iii) does not occur for oblate particles (an example
was added to the inset of Fig. \ref{F2}); and (iv) exhibits a
resonant dependence on $\tau_\theta=2/D_\theta$ (Fig. \ref{F3},
inset). For larger $F$ the mobility turns positive, reaches a
maximum, and finally decays to zero faster than $1/F$ ({\it negative
differential mobility} branch).

We now qualitatively interpret these properties by having recourse to
the analytical arguments of Refs. \cite{septate3}. First of all, the
ANM mechanism can be explained by recalling that a driven prolate
particle gets pressed longitudinally against the side walls of the
channel compartments. As in Figs. \ref{F2} and \ref{F3}, $\Delta
<2a$, the particle can slip through the wall openings only by
rotating with $\vec{v}_0$ (almost) orthogonal to the channel
cross-section \cite{ANMshape}. Again, this is a noise-activated
mechanism with Arrhenius factor $\exp(-|a-b|(F\pm v_0)/D_0)$, the
sign $\pm$ denoting the parallel or anti-parallel orientation of
$\vec {v}_0$ with respect to $\vec{F}$. Therefore, pore crossings are
more likely to take place on the left side of the compartment than on
its right side; hence the observed negative mobility. This argument
also leads to the conclusion that ANM is ruled out for prolate JPs
with $F\geq v_0$ and oblate JPs at any drive. This standstill at zero
current crossing with $\mu(F)=0$ for a prolate microswimmers shifts
to lower $F$ on increasing $D_0$, as dispersion induced by thermal
fluctuations tends to hamper the particle drift opposite to the
drive. An oblate microswimmer, oriented with its major axis
orthogonal to the compartment wall tends to propel itself parallel to
(rather than against) it, whereas the drive pulls it back to the
right (see sketch in Fig. \ref{F3}). This state of affairs cannot be
overturned by thermal noise. Thus, for an oblate JP pore-crossings to
the right are favored and the mobility is always positive.

For large drives, $F\gg v_0$, the particle mobility is suppressed.
This is a typical effect characteristic of septate channels
\cite{septate2}, which is not observed in smoothly-corrugated
channels \cite{ChemPhysChem}. The particle leaves the compartment to
the right, thus advancing a length $x_L$, only after an activation
time, $\tau_{\perp} \propto \exp(|a-b|(F+v_0)/D_0)$, it needs to line
up with the pore axis. Accordingly, $\mu \sim x_L/(2\tau_{\perp}F)$,
in fairly good agreement with our data. Finally, the inset of Fig.
\ref{F3} clearly indicates that the ANM is maximal for an optimal
choice of $\tau_\theta=2/D_\theta$ (or $l_\theta$). Note that for
$\tau_\theta \to 0$ the active random motion of a JP boils down to a
standard Brownian motion with Einstein bulk diffusivity $D_{\rm
eff}$, and, therefore, its mobility must be positive. On the other
hand, an elongated particle slips trough a pore when, after turning
perpendicular to the compartment wall, it diffuses toward the pore in
a time, $\tau_{\|}\sim y_l^2/2D_0$,  no longer than its orientation
time-constant $\tau_\theta$. Based on our data, an optimal ANM at
constant $F$, with $F<v_0$, is thus achieved by setting
$\tau_{\|}\sim \tau_\theta$.

We mention in passing that we detected ANM also by driving prolate
JPs along sinusoidally-corrugated channels (not shown). Contrary to
what happens in septate channels, negative mobility sets on for more
elongated particles, i.e., at higher $a/b$, and $\mu$ approaches
unity in the limit of large drives, $F\gg v_0$.

\section{Giant absolute negative mobility}

The magnitude of the backward rectification flow  also depends on the
compartment geometry and, in particular, on the pore size $\Delta$.
In the inset of Fig. \ref{F2} we plotted the mobility of a prolate JP
versus $\Delta$ for $F<v_0$ and different noise levels $D_0$. The
modulus of the negative mobility $|\mu|$ grows with $\Delta$ up to an
optimal value $\Delta_M$, which decreases with raising $D_0$; for
$\Delta>\Delta_M$ the mobility jumps abruptly to its bulk value,
$\mu=1$. This behavior is surprising as the ANM mechanism relies on
the blocking action of the compartment walls \cite{ANM,ANM2}. In the
case of an elongated particle such an action is strongest when
$\vec{F}$ and $\vec{v}_0$ point in the same direction
\cite{ANMshape}, which causes a rectification flow against $\vec{F}$,
i.e., negative mobility. However, as the pore size grows larger than
the particle length, $\Delta \gtrsim 2a$, one would expect ANM to
vanish. {\it Our data prove exactly the opposite}. More remarkably,
on increasing $\Delta$, $|\mu|$ overshoots to its maximum value
consistent with the ANM mechanism, that is, $|\mu_M^{\rm
ANM}|=(v_0-F)/(2F)$. This estimate of $|\mu^{\rm ANM}_M|$ refers to
the ideal optimistic case when a noiseless driven particle exits
unhindered the channel compartments to the left with speed $v_0-F$
for half of the time, whereas during the rest of the time it sits
stuck against the right compartment walls. Note that
spontaneous-direction reversals occur on the time scale
$\tau_\theta$. Most remarkably, the low-noise curves $\mu(\Delta)$ in
the inset of Fig. \ref{F2} exhibit sharp minima with
$|\mu_M=\mu(\Delta_M)|$ about twice the ANM estimate $|\mu_M^{\rm
ANM}|$.

A simple explanation of this finding is illustrated by the sketch
inserted also in Fig. \ref{F2}. The large values of $|\mu_M|$ we
observed imply that the {\it self-propulsion velocity of the driven
JP (almost) always points against $\vec{F}$}. For a particle confined
to a narrow channel with $l_\theta \gg y_L$, this is certainly true
for half the time. For the remaining half of the time, however,
$\vec{v}_0$ would spontaneously orient itself parallel to $\vec{F}$.
This means that the collisions with the walls (and not the
fluctuations of $\theta(t)$) must be responsible for the quick
rotation of $\vec{v}_0$ to the opposite direction. For large $\Delta$
the compartment walls shrink to a pair of perpendicular winglets of
length $\delta=(y_L-\Delta)/2$, sticking out of the channel walls. An
elongated microswimmer hitting one such obstacle from the left,
tumbles over it under the action of the torque exerted by the drive.

Of course, such a tumbling mechanism is most effective under a few
simple conditions: (i) the tumbling time, $\tau_\delta\sim \pi
I/Fa$, with $I=(a^2+b^2)/4$ denoting the moment of inertia of an
elliptical disk of unit mass, has to be much shorter than
$\tau_\theta$; (ii) $\delta$ has to be the shortest possible, to
minimize the blocking action of the compartment walls; (iii) the
noise level has to be lowered as much as possible, lest the
particle diffuses past the obstacle without tumbling. An optimal
choice for $\delta$ seems $\delta_M\sim b/2$, which corresponds to
$\Delta_M\sim y_L-b$ (see Fig. \ref{F2}, inset). For shorter
$\delta$ the particle can slide over the obstacle without
rotating. Under these conditions $\mu_M \simeq -(v_0-F)/F$ and its
modulus can grow very large, indeed! We propose to term this
phenomenon {\it giant negative mobility} (GNM).

\begin{figure}
\centering
\includegraphics[width=0.45\textwidth]{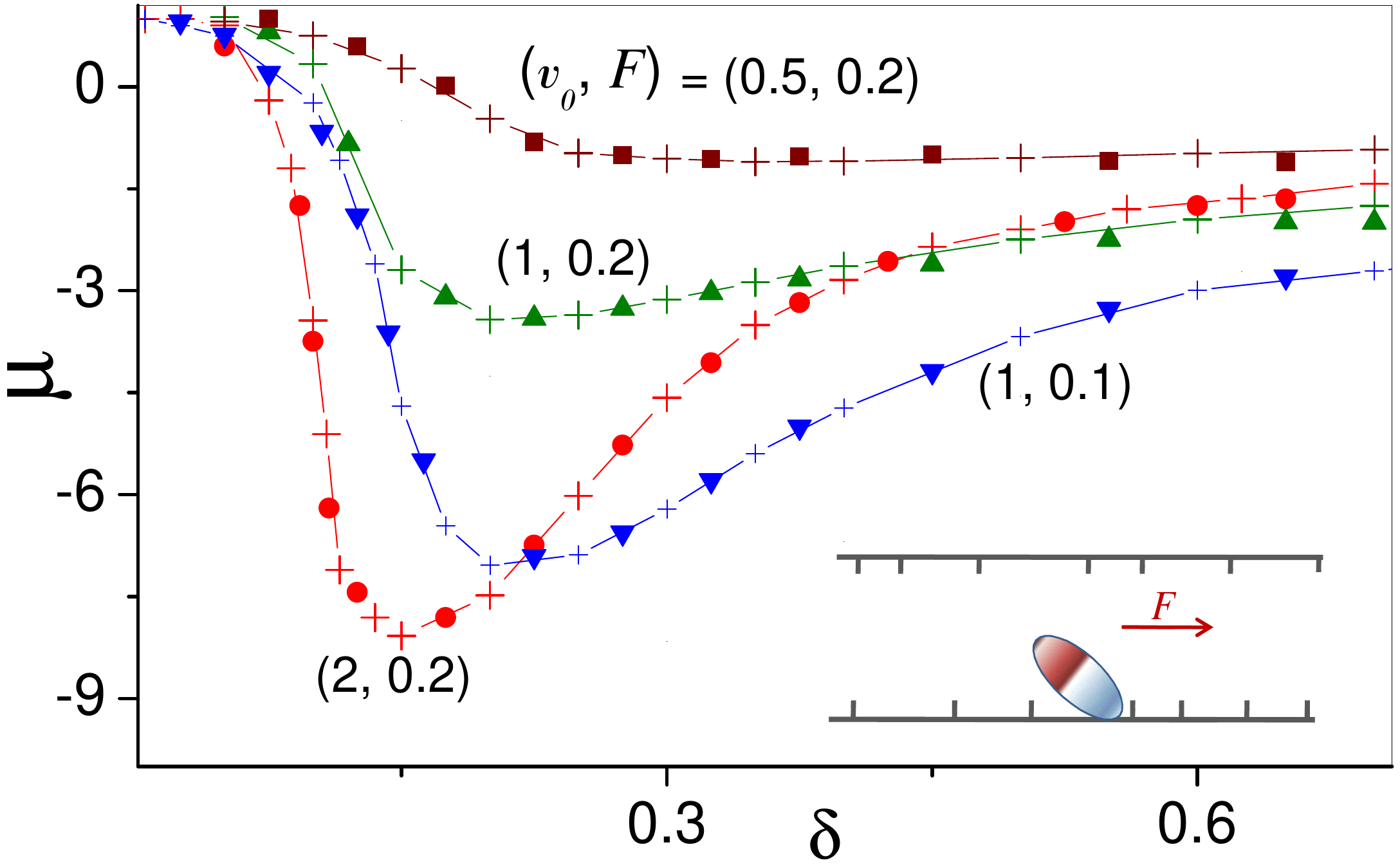}
\caption {(Color online) Giant negative mobility for a prolate JP
driven along a rough channel of width $y_L=10$ with different
$v_0$ and $F$ (see legend). On both sides winglets of length
$\delta$ are separated by a random distance $l_n$, uniformly
distributed in the interval $[5, 15]$ (solid symbols, see text);
for a comparison we also simulated the corresponding case of
equally-spaced winglets with $l_n=10$ (crosses). Other simulation
parameters are: $a=1$, $b=0.25$, $D_0=0.02$, and $D_\theta=6\times
10^{-3}$.} \label{F4}
\end{figure}

GNM is sustained by the self-propulsion of the driven microswimmers
themselves, the drive just helping orient their velocities opposite
to it. In such a scheme, contrary to ANM \cite{ANM}, the obstacles
fabricated along the channel walls do not act as traps, but rather as
centers of rotation (tumbling). To better assess the robustness of
GNM, in Fig. \ref{F4} we plotted $\mu$ versus $\delta$ for both
periodically and randomly distributed winglets, all of the same
length $\delta$. Randomized winglet distributions have been generated
by scrambling regular distributions with period $x_L$, that is, by
spacing the winglets a distance $l_n$ apart, with
$l_n=x_{n+1}-x_n=x_L(1+\delta_n)$, $x_n$ being the position of the
n-th winglet and $\delta_n$ a random number uniformly distributed in
the interval $[-0.5,0.5]$. To closer mimic wall roughness, the
winglet distributions on the upper and lower channel walls were
independently generated. We selected particle and compartment sizes
consistent with the experimental setups described in the literature
\cite{Bechinger}. For both winglet distributions our estimates for
$\Delta_M$ and $\mu_M$ work quite well, meaning that GNM ought to be
experimentally detectable at room temperature.

For all curves displayed in Fig. \ref{F4}, $|\mu_M|$ comes quite
close to its upper bound $2|\mu_M^{\rm ANM}|$. As anticipated, on
lowering the self-propulsion diffusivity, $v_0^2/2D_\theta$,
against $D_0$, the optimal winglet length $\delta_M$ shifts to
higher values and $|\mu_M|$ slowly diminishes. Moreover, since
particle tumbling is a local mechanism, GNM is rather insensitive
to the actual distribution of the channel winglets. Analogously,
winglets can be replaced by protrusions of different geometry to
better model the roughness of channel walls. As long as they are
sharp enough to engage the tips of rod-like JPs, the GNM mechanism
does work (possibly facilitated by the hydrodynamic attraction
between microswimmers and channel walls); if not, the smooth
channel situation discussed at the very beginning would be
recovered.

To emphasize the mechanisms responsible for the negative mobility of
elongated JP's, in Eq. (\ref{LE}) we neglected a number of additional
effects which may prove experimentally appreciable. We started
assuming that self-propulsion is fully described by the random
velocity ${\vec v}_0$, thus implying the absence of chiral (or
torque) and inertial (or viscous) terms \cite{Lowen}. We also adopted
scalar translational, $D_0$, and orientational diffusivity, $D_c$,
despite the elongated shape of the active swimmer. A more detailed
modeling of the active roto-translational dynamics \cite{Lowen_basic}
would have no substantial impact on the occurrence of negative
mobility; the corresponding hydrodynamic corrections to the particle
dynamics can be accounted by an appropriate rescaling of the model
parameters \cite{Dhar}. More importantly, we ignored hydrodynamic
effects, which not only favor clustering in dense mixtures of JP's
\cite{Ripoll,Buttinoni}, but may even cause their capture by the
channel walls \cite{Takagi}. Therefore, in the case of single
diffusing JP's, hydrodynamic effects can only reinforce the flow of
particles at the boundary, thus enhancing the predicted diffusion
transients and the tumbling mechanism responsible for GNM. On the
contrary, hydrodynamic effects are known to impact on the selective
translocation of elongated JP's through narrow pores \cite{Ai},
$\Delta \gg y_L$, and, therefore, are likely to hamper the direct
observation of ANM.

\section{Conclusions}

We numerically simulated the transport of elongated Janus particles
driven along a narrow channel. Key transport quantifiers, like
mobility and diffusivity, strongly depend on the particle shape.
Diffusion in smooth channels is characterized by exceedingly long
transients, either ballistic or non-diffusive, respectively, for
prolate and oblate active microswimmers. In compartmentalized
channels with narrow pores prolate Janus particles undergo absolute
negative mobility, as an effect of the translational symmetry
breaking due to the drive. More importantly, when the compartment
dividers shrink to small side winglets, possibly randomly distributed
along the channel walls, rod-like active particles greatly enhance
their negative mobility, as the combined action of drive and channel
roughness systematically reorients the particle self-propulsion
velocity opposite to the drive itself. As the geometric and dynamical
parameters used in our simulations closely compare with those
reported for actual experimental set-ups, we are confident that giant
negative mobility can soon be demonstrated, thus allowing a more
effective transport control of active microswimmers. Such a control
technique can be exploited, e.g., for medical applications, such as
drug delivery via JPs to physiological opposing regions. Moreover, a
dilute binary mixture of JP's of different shape can be driven along
one such stylized rough channel so as to generate a two-way traffic
(i.e., with mobilities of opposite sign), without having recourse to
a parallel two-channel architecture with opposite drives..

\section*{Acknowledgements} We thank RICC for computational resources.
P.H. acknowledges support from the cluster of excellence Nanosystems
Initiative Munich (NIM). F.M. acknowledges support by the program for
internationalization of the Augsburg University. FN is partially
supported by the RIKEN iTHES Project, MURI Center for Dynamic
Magneto-Optics, JSPS-RFBR contract no. 12-02-92100, Grant-in-Aid for
Scientific Research (S), MEXT Kakenhi on Quantum Cybernetics and the
JSPS via its FIRST program.

\end{document}